\documentclass[12pt, leqno]{article}

\usepackage[letterpaper,top=2cm,bottom=2cm,left=3cm,right=3cm, marginparwidth=2cm]{geometry}

\usepackage{amsthm}
\usepackage{graphicx}
\usepackage{microtype}
\usepackage[usenames,dvipsnames]{xcolor}
\usepackage{amsmath,amssymb,amsfonts,latexsym,cancel}
\usepackage{amsmath, amstext, amsfonts}
\usepackage{float}
\usepackage{caption}
\usepackage{appendix}
\usepackage{multirow}

\usepackage{subfig}
\usepackage{tabularx}
\usepackage{algorithm}
\usepackage[noend]{algpseudocode}

 \usepackage{mathptmx}      

\newtheorem{theorem}{Theorem}[section]

\newtheorem{definition}[theorem]{Definition}

\begin{document}

{\centering \LARGE A novel encryption algorithm using multiple semifield S-boxes based on permutation of symmetric group\par}
\bigskip

{\centering \normalsize Iqtadar Hussain$^1$, Amir Anees$^2$, Temadher Alassiry Al-Maadeed$^1$, M. T. Mustafa$^1$ \par}
{\centering\itshape $^{1}$Department of Mathematics, Statistics and Physics, Qatar University, Doha, 2713, State of Qatar \par}
{\centering\itshape $^{2}$Department of Computer Science and Information Technology, La Trobe University, Melbourne, Australia \par}
{\centering\itshape   iqtadarqau@qu.edu.qa, a.anees@latrobe.edu.au, t.alassiry@qu.edu.qa, tahir.mustafa@qu.edu.qa   \par}

\begin{abstract}
With the tremendous benefits of internet and advanced communications, there is a serious threat from the data security perspective. There is a need of secure and robust encryption algorithm that can be implemented on each and diverse software and hardware platforms. Also, in block symmetric encryption algorithms, substitution boxes are the most vital part. In this paper, we investigate semifield substitution boxes using permutation of symmetric group on a set of size 8 \(S_8\) and establish an effective procedure for generating \(S_8\) semifield substitution boxes having same algebraic properties. Further, the strength analysis of the generated substitution boxes is carried out using the well-known standards namely bijectivity, nonlinearity, strict avalanche criterion, bit independence criterion, XOR table and differential invariant. Based on the analysis results, it is shown that the cryptographic strength of generated substitution boxes is on par with the best known $8\times 8$ substitution boxes. As application, an encryption algorithm is proposed that can be employed to strengthen any kind of secure communication. The presented algorithm is mainly based on the Shannon idea of substitution-permutation (S-P) network where the process of substitution is performed by the proposed \(S_8\) semifield substitution boxes and permutation operation is performed by the binary cyclic shift of substitution box transformed data. In addition, the proposed encryption algorithm utilizes two different chaotic maps. In order to ensure the appropriate utilization of these chaotic maps, we carry out in-depth analyses of their behavior in the context of secure communication and apply the pseudo-random sequences of chaotic maps in the proposed image encryption algorithm accordingly. The statistical and simulation results imply that our encryption scheme is secure against different attacks and can resist linear and differential cryptanalysis. \\

\end{abstract}

\section{Introduction}
\label{intro}
Substitution box (S-box) is the most important element in any block cipher. Being the only nonlinear component of block cipher, a substitution box plays the vital role of inducing the non-linearity between the ciphertext and the secret key in block cipher encryption algorithms. Hence the S-box  \cite{ref_8,ref_2,ref_9,hussain_1} is used as a tool in block ciphers  to enhance their resistance against differential and linear cryptanalysis \cite{ref_1}. The S-box of Advanced Encryption Standard (AES) \cite{aes} can be defined via the arrangement of eight Boolean functions with the domain and the range of each function as $GF(2^8)$ and \(GF(2)\) respectively. In \cite{iqi_1}, Rafael Alvarez and Gary McGuire have provided some measures to analyze the cryptographic potency of S-box. These measures determine the cryptographic invulnerability of S-boxes against special categories of promising attacks. In literature, it can be seen that along with injective and surjective properties of S-boxes, only $S_8$ S-boxes proposed by Iqtadar et. al., \cite{iqi_2}, and AES S-box have high-quality values close to optimal readings for all of these criteria. The reason to get similar analysis of these boxes is that $S_8$ S-boxes are permutation extension of AES S-box based on a symmetric group. A large number of papers have been devoted to studying finite field S-boxes and their applications. The reader is referred to
\cite{ref_2,ref_9,hussain_1,aes,iqi_1} for the literature on finite field S-boxes and their applications.
\subsection{Algebraic structure of semifield}
\label{sec:1}
This paper investigates S-boxes over the algebraic structure of semifield, defined as follows.
\begin{definition}
A finite semifield consists of a finite set $S$ with two binary operations ``+" and ``$\cdot$" such that the following axioms hold:
\begin{itemize}
	\item ($S$,+) forms a group with identity element $0$.
	\item $\forall a,b \in S$: If $ab=0$ then $a=0$ or $b=0$.
	\item $\forall a,b \in S$: $a(b+c)=ab+ac$ and $(a+b)c=ac+bc$.
	\item $\forall a \in S$ $\exists$ a neutral element for $\times$ denoted as $e$ which satisfies: $ea=ae=a$.
\end{itemize}
\end{definition}
For example the following system $S$ is a proper semifield having order 16. Let us consider $GF(2^2)=\left\{ 0, 1, \omega, \omega^2   \right\}$ as a base field $F$ for the construction of the semifield $S_{2^4}$. The form of the elements of semifield is $u$+$\lambda$$v$, where $u$, $v$ $\in$ $F$. The addition in the semifield $S_{2^4}$ is defined using the addition of field $F$ as
$
(u+\lambda v)+(x+\lambda y)=(u+v)+\lambda(x+y).
$
Multiplication in the semifield $S_{2^4}$ is defined in terms of the multiplication and addition of $F$, in the following manner
$$
(u+\lambda v)(x+\lambda y)=(ux+v^2y)+\lambda(vx+u^2y+v^2y^2).
$$

\begin{table}[h!]
\centering 
\begin{tabular}{c | c c c c} 
\hline\hline
Rows/Columns & 0 & 1 & $\omega$ & $\omega^2$= 1+$\omega$ \\
\hline\hline 
inserts single horizontal line
0&0&$\lambda$&$\lambda$$\omega$&$\lambda$$\omega^2$\\
1&1&1+$\lambda$&1+$\lambda$$\omega$&1+$\lambda$$\omega^2$\\
$\omega$&$\omega$&$\omega$+$\lambda$&$\omega$+$\lambda$$\omega$&$\omega$+$\lambda$$\omega^2$\\
$\omega^2$= $1$+$\omega$&$\omega^2$&$\omega$+$\lambda$&$\omega^2$+$\lambda$$\omega$&$\omega^2$+$\lambda$$\omega^2$\\
\hline 
\end{tabular}
\caption{ Semifields $S_{2^4}$ with 16 elements}
\label{table_iqi_1} 
\end{table}

In semifield, there is no associative property as contrary to let say Galois Field. This characteristic of having no associative property makes S-boxes way more difficult to break using linear or differential cryptanalysis. The known S-boxes based on finite fields are good enough against known algebraic attacks, yet there could be some unknown algebraic attacks that may weaken the confusion creating capability of finite field S-boxes.  To overcome this possible weakness, Dumas- Orfila \cite{iqi_3} proposed to enhance the algebraic strength of S-boxes by considering finite semifields instead of finite fields to generate S-Boxes.
The known classification of semifields is up to $2^6$, therefore, it is a challenge to use the small structure $2^4$ for the construction of $2^8$ S-boxes. Dumas- Orfila \cite{iqi_3} utilized pseudo-extensions of semifield of order $2^4$ to construct 12781 non-equivalent S-Boxes with maximum  cryptographic properties. In this paper, corresponding to each and every non-equivalent S-box of Dumas- Orfila \cite{iqi_3}, we generate $8!$ S-boxes with same cryptographic properties . This is achieved by the action of symmetric group of permutation on every non-equivalent S-box of \cite{iqi_3}. Moreover, the non-equivalent S-boxes will behave like a set. Because all 12,781 S-boxes of \cite{iqi_3} are non-equivalent, therefore, we have 12,781 mutually disjoint classes of S-boxes. Further, we have shown the application of these S-boxes in image encryption.

The paper is organized as follows. The necessary details about the chaotic maps employed in this work, along with their relevance to secure communication, are provided in Section 2. Section 3 is divided in two parts, giving the construction methodology of proposed S-boxes followed by analyses for evaluating their strength. Section 4 presents the proposed block cipher. The experimental and security analysis for the proposed cipher are presented in Section 5 and Section 6, respectively. Finally,  Section 7 gives the conclusion.

\section{Chaos and their significance in secure communication}

In recent years, number of articles published on chaotic security \cite{ref_10} have proven to be weak against well known attacks \cite{ref_3,ref_4}. This is due to the fact that the researchers did not analyze chaotic phenomena in detail with respect to secure communication \cite{ref_5,ref_6}. Although there are many similarities between chaotic behavior of chaotic maps and randomness required in security \cite{ref_7} but it still needs proper attention to be applied in this field.

In this paper, we analyze a modified version of logistic chaotic map \cite{log} in detail with respect to secure communication and observe the regions of interest which can be applicable in communication security. We have employed two chaotic maps in our proposed encryption algorithm; modified logistic map and Tangent Delay Ellipse Reflecting Cavity Map System (TD-ERCS) \cite{tdercs}. The modified logistic chaotic map employed in our work is given as \cite{ex_log}:

\begin{equation}
			x_{n} = rx_{n-1}(1-x_{n-1}^{1-b}).
\end{equation}
\begin{figure*}
  \centering
		\subfloat[]{\includegraphics[width=0.33\textwidth]{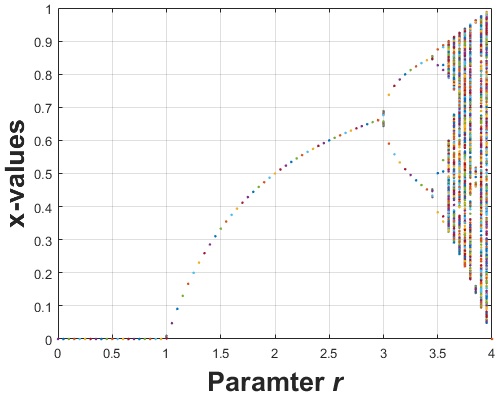} }
	  \subfloat[]{\includegraphics[width=0.33\textwidth]{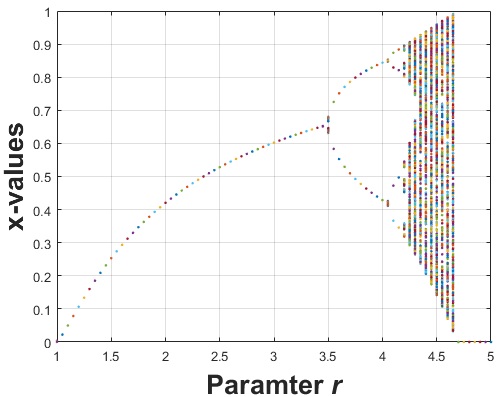} }
		\subfloat[]{\includegraphics[width=0.33\textwidth]{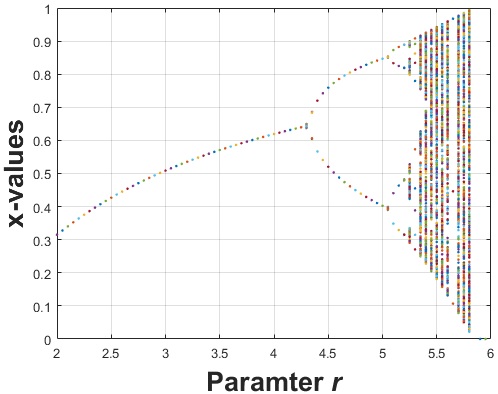} }\\
		\subfloat[]{\includegraphics[width=0.33\textwidth]{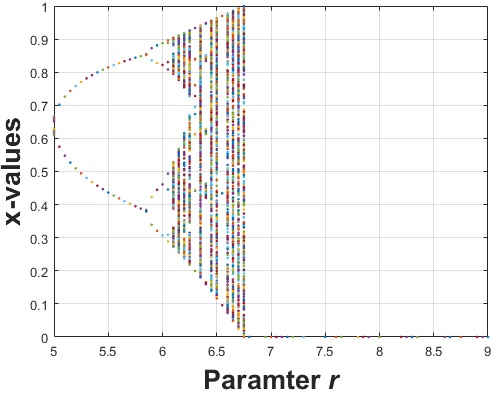} }
	  \subfloat[]{\includegraphics[width=0.33\textwidth]{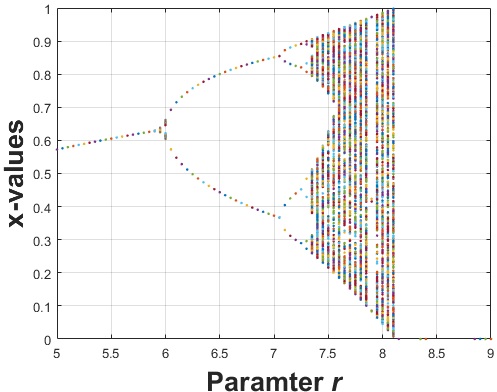} }
		\subfloat[]{\includegraphics[width=0.33\textwidth]{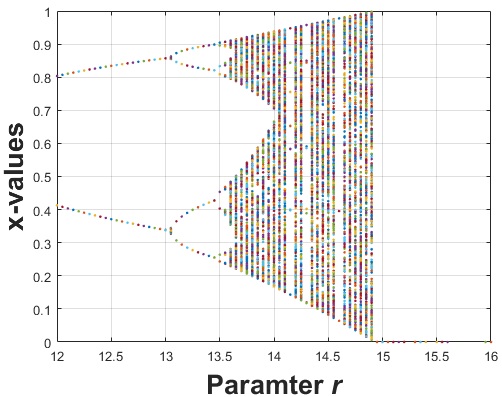} }
  \caption{Bifurcation diagram of modified logistic map showing the plots of all the \(x\)-vectors graphed against \(r\) for different values of \(b\). In order to observe the long-term impact of modified logistic map, the values of \(x\) are plotted after 500 iterations.   (a) \(b\) = 0, (b) \(b\) = 0.2, (c) \(b\) = 0.4, (d) \(b\) = 0.5, (e) \(b\) = 0.6 and (f) \(b\) = 0.8}
	\label{bif}
\end{figure*}
where, \(x_0 \in (0,1)\), \(b \in [0,1)\) and \(r\) (depending on \(b\))  are initial seed parameters. At \(b = 0\), the above equation becomes the logistic map. At a specific value of \(b\), \(r\) is a control parameter which specifies the overall behavior of this chaotic map. In physical terms, it can be defined as the heating in a convention or the growth rate of population of a typical bacteria type etc. Fig.\hspace{1mm}\ref{bif} presents the bifurcation diagram of modified logistic map which is a combined presentation of all the graphs plotted \(x\) against \(r\) for different values of \(b\). Fig.\hspace{1mm}\ref{bif}(a) shows the bifurcation diagram at \(b = 0\) which is equivalent of the bifurcation diagram of logistic map. The range of the randomness in the bifurcation diagrams at all the values of \(b\) is almost same; the only difference between these figures is the shifting of the range for the values of \(r\) which is a significant advantage as compared to logistic map.

\subsection{Chaotic range analysis}

Taking the case \(b = 0.2\) (Fig.\hspace{1mm}\ref{bif}(b)) we analyze the ranges which are \(r\) dependable. Based on Fig.\hspace{1mm}\ref{bif}(b), we can divide the interval for \(r\) into three segments:

\begin{itemize}
	\item When \(r \in [1, 3.5]\), the iteration sequence for \(x\) attains a constant value past some iterations displaying a stable behavior. Fig.\hspace{1mm}\ref{bif_a}(a) shows the bifurcation diagram of modified logistic map for \(r = 1\) to \(r = 3.5\), Fig.\hspace{1mm}\ref{bif_a}(b) demonstrates the iteration sequence for initial conditions  \(r = 2.5\), \(b = 0.2\), \(x_0 = 0.5\) and Fig.\hspace{1mm}\ref{bif_a}(c) illustrates the iteration sequence for initial conditions  \(r = 3.2\), \(b = 0.2\), \(x_0 = 0.5\). It should be noted that the iteration sequences converge to a stable point beyond some iterations and hence are not suitable for utilization in secure communication.
	\begin{figure}[h]
 \centering
		\subfloat[]{\includegraphics[width=0.33\textwidth]{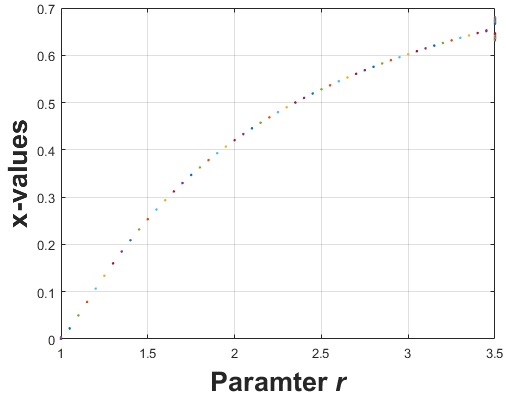} }
	  \subfloat[]{\includegraphics[width=0.33\textwidth]{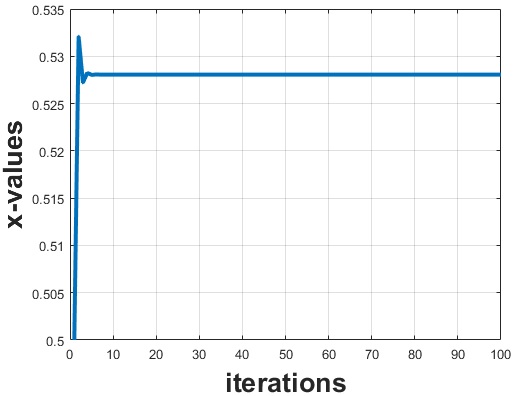} }
		\subfloat[]{\includegraphics[width=0.33\textwidth]{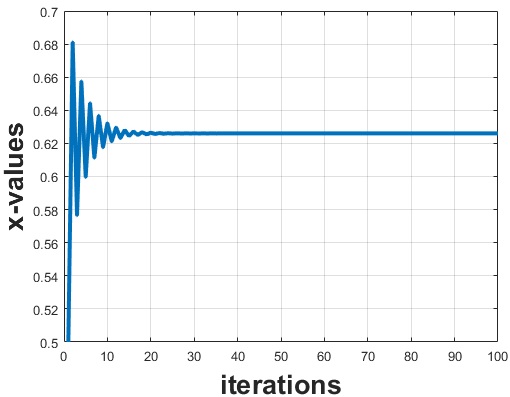} }
  \caption{(a).  Bifurcation diagram of modified logistic map \(x\) plotted against \(r = 1\) to \(r = 3.5\) for \(b = 0.2\). (b) and (c). The iteration sequence plotted respectively for the initial conditions (i) \(r = 2.5\), \(b = 0.2\), \(x_0 = 0.5\) and (ii) \(r = 3.2\), \(b = 0.2\), \(x_0 = 0.5\).}
	\label{bif_a}
\end{figure}
	\item When \(r \in (3.5, 4.2)\), the iteration sequence for \(x\) exhibits periodic behavior with different periodicity after some iterations. Fig.\hspace{1mm}\ref{bif_b}(a) illustrates the bifurcation diagram of modified logistic map for \(r = 3.5\) to \(r = 4.2\), Fig.\hspace{1mm}\ref{bif_b}(b) demonstrates the iteration sequence for initial conditions  \(r = 3.7\), \(b = 0.2\), \(x_0 = 0.5\) and Fig.\hspace{1mm}\ref{bif_b}(c) displays the iteration sequence for initial conditions  \(r = 4.1\), \(b = 0.2\), \(x_0 = 0.5\). It is worth noticing that the iteration sequences exhibit periodic behavior after some iterations having periodicity 2 for Fig.\hspace{1mm}\ref{bif_b}(b) and 4 for Fig.\hspace{1mm}\ref{bif_b}(c) and thus are not suitable for usage in secure communication.
	\begin{figure}[h]
  \centering
	\subfloat[]{\includegraphics[width=0.33\textwidth]{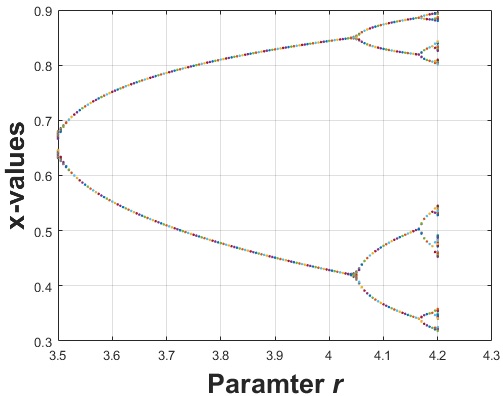} }
	  \subfloat[]{\includegraphics[width=0.33\textwidth]{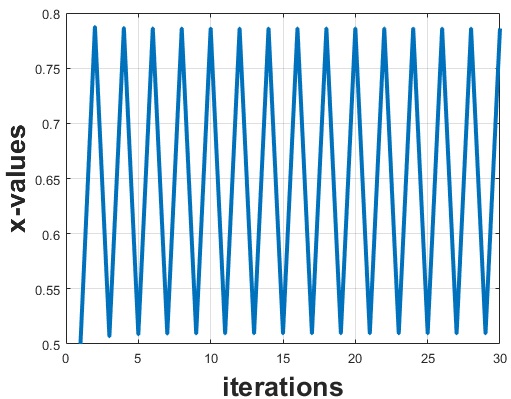} }
		\subfloat[]{\includegraphics[width=0.33\textwidth]{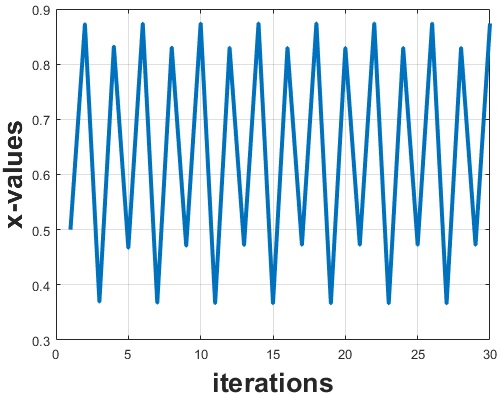} }
  \caption{(a). Bifurcation diagram of modified logistic map \(x\) plotted against \(r = 3.5\) to \(r = 4.2\) for \(b = 0.2\). \\
	(b) and (c).  The iteration sequence plotted respectively for the initial conditions (i)  \(r = 3.7\), \(b = 0.2\), \(x_0 = 0.5\) and (ii)  \(r = 4.1\), \(b = 0.2\), \(x_0 = 0.5\).}
	\label{bif_b}
\end{figure}
	\item Whilst \(r \in [4.2, 4.6]\), the iteration sequence for \(x\) exhibits chaotic behavior. Fig.\hspace{1mm}\ref{bif_c}(a) illustrates the bifurcation diagram of modified logistic map for \(r = 4.2\) to \(r = 4.6\), Fig.\hspace{1mm}\ref{bif_c}(b) demonstrates the iteration sequence for initial conditions  \(r = 4.3\), \(b = 0.2\), \(x_0 = 0.5\), Fig.\hspace{1mm}\ref{bif_c}(c) illustrates the iteration sequence for initial conditions  \(r = 4.5\), \(b = 0.2\), \(x_0 = 0.5\) and Fig.\hspace{1mm}\ref{bif_c}(d) illustrates the iteration sequence for initial conditions  \(r = 4.56\), \(b = 0.2\), \(x_0 = 0.5\). It should be noted that the iteration sequences are completely random illustrating a chaotic behavior and thus can be used in secure communication. However, the whole range between \(r = 4.2\) to \(r = 4.6\) does not have a chaotic illustration. For example, when \(r = 4.41\), it shows periodic behavior and thus is not suitable to be applicable in secure communication as can be seen in Fig.\hspace{1mm}\ref{bif_c}(e) displaying the iteration sequence for initial conditions of \(r = 4.41\), \(b = 0.2\), \(x_0 = 0.5\). Furthermore, at \(r = 4.52\), it shows periodic behavior and thus not suitable to be applicable in secure communication as can be seen in Fig.\hspace{1mm}\ref{bif_c}(f) that shows the iteration sequence for initial conditions of \(r = 4.52\), \(b = 0.2\), \(x_0 = 0.5\).
	
	\begin{figure}[h!]
  \centering
		\subfloat[]{\includegraphics[width=0.33\textwidth]{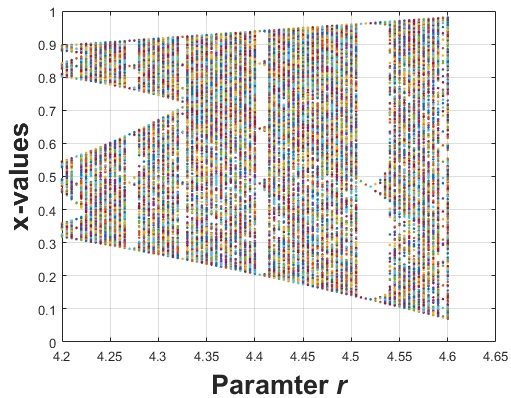} }
	 \subfloat[]{\includegraphics[width=0.33\textwidth]{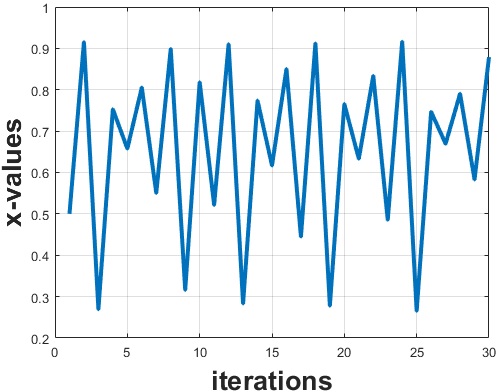} }
		\subfloat[]{\includegraphics[width=0.33\textwidth]{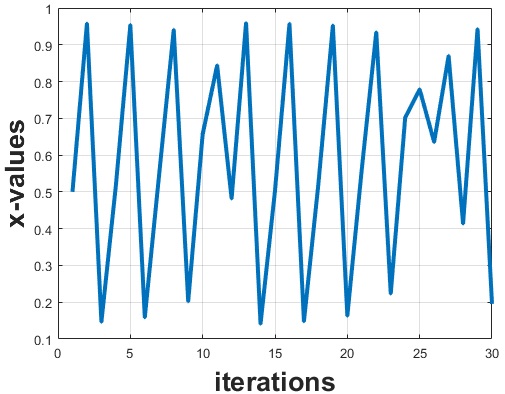} } \\
		\subfloat[]{\includegraphics[width=0.33\textwidth]{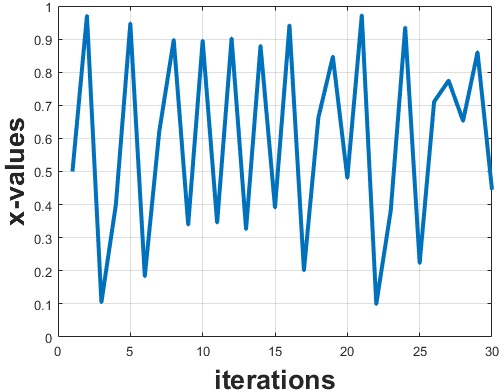} }
	  \subfloat[]{\includegraphics[width=0.33\textwidth]{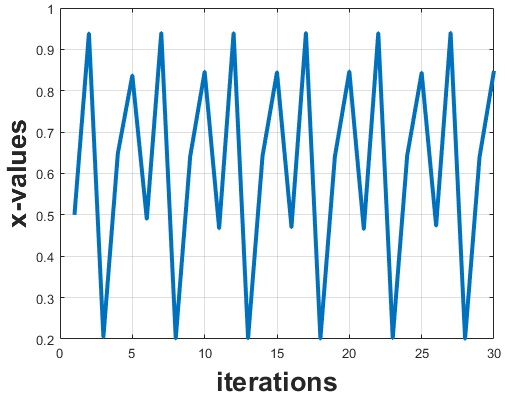} }
		\subfloat[]{\includegraphics[width=0.33\textwidth]{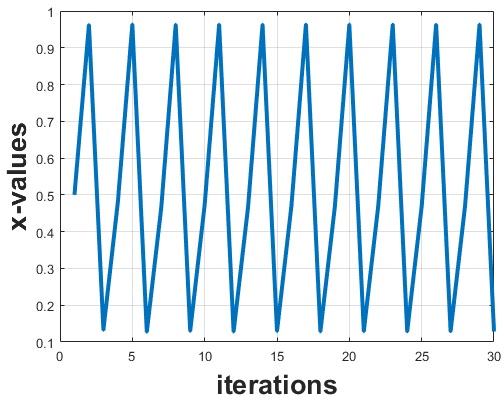} }
  \caption{(a). Bifurcation diagram of modified logistic map \(x\) plotted against \(r = 4.2\) to \(r = 4.6\) for \(b = 0.2\). \\
	(b), (c), (d), (e) and (f). The iteration sequence plotted respectively for initial conditions (i) \(r = 4.3\), \(b = 0.2\), \(x_0 = 0.5\), (ii)  \(r = 4.5\), \(b = 0.2\), \(x_0 = 0.5\),  (iii)  \(r = 4.56\), \(b = 0.2\), \(x_0 = 0.5\),  (iv)  \(r = 4.41\), \(b = 0.2\), \(x_0 = 0.5\) and (v) \(r = 4.52\), \(b = 0.2\), \(x_0 = 0.5\).}
	\label{bif_c}
\end{figure}

\end{itemize}

\subsection{Randomness analysis}
Similar to above, we can also analyze the ranges of bifurcation diagrams for the other values of \(b\) and can observe their chaotic ranges. Apart from visually analyzing the randomness of these ranges, we have tested the sequence using the NIST statistical analysis suite \cite{nist} which contains 15 statistical tests. Although these tests were primarily designed for a larger stream of binary bits (usually $>$ 10000), these tests can also be used on shorter streams as well as on integer values. The results of these tests on chaotic sequences are listed in Table\hspace{1mm}\ref{t_1} showing that the generated chaotic sequences pass the randomness tests.

\begin{table}[h]
\centering
\label{t_1}       
\begin{tabular}{lll}
\hline\noalign{\smallskip}
Statistical Test & $P$ Value & Decision  \\
\noalign{\smallskip}\hline\noalign{\smallskip}
Frequency (Mono Bit) Test  & 0.9941 & Passed \\
(Blocks sizes: 3, 4, 5, 6, 7, 8)  & 0.6724 & Passed \\
The Runs Test											 	& 0.0251 & Passed \\
Tests for the Longest-Run-of-Ones in a Block \\			
(Blocks size: 8)	                                                                       & 0.4521 & Passed \\
The Binary Matrix Rank Test					\\												
(4 Matrices, Rows: 8, Columns: 8)   			& 0.0745 & Passed \\
 (16 Matrices, Rows: 4, Columns: 4)				& 0.1214 & Passed \\
The Discrete Fourier Transform (Spectral) Test					& 0.1231 & Passed \\
The Non-overlapping Template Matching Test                        & 0.0254 & Passed \\
 (Template length = 4, Blocks = 2, 4, 8) 	\\
The Overlapping Template Matching Test \\			
(Template length = 4, Blocks = 4, 8) 			& 0.5987 & Passed \\
Maurer's Universal Statistical Test											& 0.2917 & Passed \\
The Approximate Entropy Test														& 0.3680 & Passed \\
The Cumulative Sums (Cusums) Test			 									& 0.4198 & Passed \\
\noalign{\smallskip}\hline
\end{tabular}
\caption{NIST statistical tests for checking the randomness of chaotic sequences obtained from modified logistic map for initial conditions  \(r = 4.3\), \(x_0 = 0.5\) and \(b = 0.2\). The findings of the tests imply that the chaotic sequences pass the randomness statistical tests}
\end{table}

\subsection{Sensitivity analysis}

\begin{figure}
  \centering
	\subfloat[]{\includegraphics[width=0.33\textwidth]{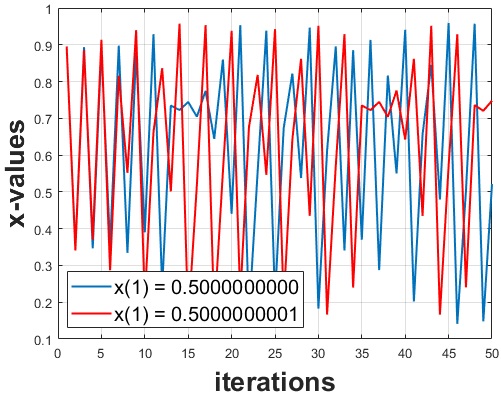} }
		\subfloat[]{\includegraphics[width=0.33\textwidth]{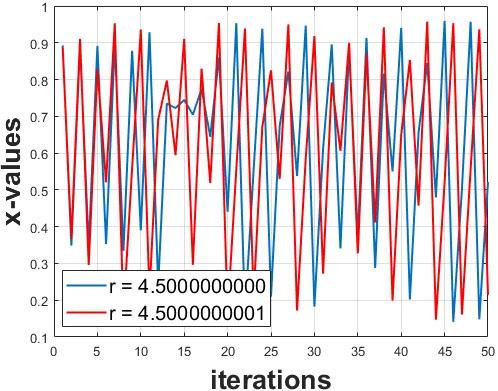} }
		\subfloat[]{\includegraphics[width=0.33\textwidth]{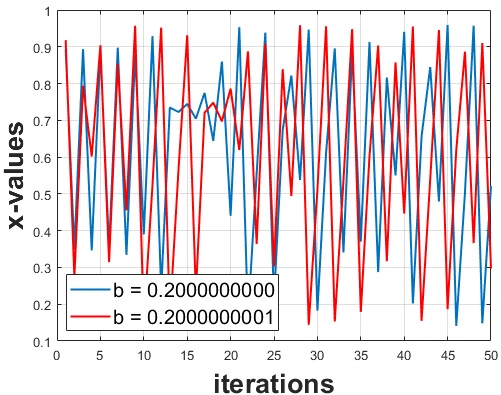} }
	  \caption{(a), (b) and (c). Chaotic sequence for modified logistic map respectively for initial conditions (i) \(r = 4.5\), \(x_0 = 0.5000000000\), \(b = 0.2\) and \(r = 4.5\), \(x_0 = 0.5000000001\), \(b = 0.2\), (ii) \(r = 4.5000000000\), \(x_0 = 0.5\), \(b = 0.2\) and \(r = 4.5000000001\), \(x_0 = 0.5\), \(b = 0.2\) and (iii)  \(r = 4.5\), \(x_0 = 0.5\), \(b = 0.2000000000\) and \(r = 4.5\), \(x_0 = 0.5\), \(b = 0.2000000001\).}
	\label{sen}
\end{figure}

Randomness alone is not enough for any chaotic sequence to be applied in secure communication \cite{stat1,stat2}. It is necessary that the chaotic sequence should be sensitive to the initial conditions. For the example taken of modified logistic map, we have considered three scenarios for sensitivity analysis; one for each initial condition of each of the parameter \(x_0\), \(r\) and \(b\).
For comparison, Fig.\hspace{1mm}\ref{sen}(a) displays the graphs of chaotic sequence for modified logistic map for initial conditions  \(r = 4.5\), \(x_0 = 0.5000000000\), \(b = 0.2\) and for slightly changed initial conditions  \(r = 4.5\), \(x_0 = 0.5000000001\), \(b = 0.2\).
It can be observed that both the sequences are similar initially for up to 15 iterations, but beyond that, both sequences become completely out of phase from each other. Moreover, the modified logistic map is sensitive to the initial conditions for \(r\) and \(b\). Fig.\hspace{1mm}\ref{sen}(b) illustrates the chaotic sequence for initial conditions  \(r = 4.5000000000\), \(x_0 = 0.5\), \(b = 0.2\) and in the same figure, a graph is plotted for initial conditions \(r = 4.5000000001\), \(x_0 = 0.5\), \(b = 0.2\). Similarly, Fig.\hspace{1mm}\ref{sen}(c) displays the chaotic sequence for initial conditions \(r = 4.5\), \(x_0 = 0.5\), \(b = 0.2000000000\) and in the same figure, a graph is plotted for initial conditions  \(r = 4.5\), \(x_0 = 0.5\), \(b = 0.2000000001\). It can be observed that at the beginning the sequences in both the figures are nearly same but with increaseed number of iterations, all the sequences are individually recognizable.
For the other chaotic map TD-ERCS, one can similarly show that it has same properties as modified logistic map. The TD-ERCS map provides  two chaotic sequences, \(x\) and \(k\), given mathematically as \cite{tdercs}:

\begin{equation}
	\begin{cases}
    x_n = -\frac{2k_{n-1}y_{n-1}+x_{n-1}(\mu^2-k_{n-1}^2)}{\mu^2+k_{n-1}^2}.\\
    k_n = -\frac{2k_{n-m}^{'}-k_{n-1}+k_{n-1}k_{n-m}^{'2}}{1+2k_{n-1}k_{n-m}^{'}-k_{n-m}^{'2}}
  \end{cases}
\end{equation}

where,

\begin{equation}
    k_n^{'} = -\frac{x_{n}}{x_{n}}\mu^2.
\end{equation}
\begin{equation}
    y_n = k_{n-1}(x_n - x_{n-1})+y_{n-1}.
\end{equation}
\begin{equation}
	k_{n-m} =
  \begin{cases}
    \frac{x_{n-1}}{y_{n-1}}\mu^2       		& \quad if\ \  n < m\\
    \frac{x_{n-m}}{y_{n-m}}\mu^2       		& \quad if\ \  n \geq m.
  \end{cases}
\end{equation}

The initial seed parameters are,

\begin{equation}
\begin{cases}
x_0\in[-1, 1],\\
			tan\alpha\in(-\infty, \infty),\\
			\mu\in(0.05, 1),\\
			m = 2, 3, ..., n.
\end{cases}
\end{equation}

These initial parameters, lead to
\begin{equation}
y_0 = \mu\sqrt{1-x_{0}^{2}}
\end{equation}

\begin{equation}
k_{0}^{'} = -\frac{x_0}{y_0}\mu^2
\end{equation}

\begin{equation}
k_{0} = \frac{tan\alpha+k_{0}^{'}}{1-k_{0}^{'}tan\alpha}
\end{equation}

\section{Proposed S-boxes}

The ideal scenario for the construction of $8\times8$ S-box based on semifield is to find a semifield with order 256. But unfortunately, the complete classification of semifield is still unknown and, in literature,  the largest known classification of semifield with characteristic 2 is of order 64. Therefore, it is not possible  to find a direct method to incorporate the nonlinear component S-box in block cipher based for this structure. But Jean et al. \cite{iqi_3} have shown an indirect way for the construction of S-box based on semifield. Our proposed construction method in this work consists of two steps.
\begin{description}
	\item[Step 1: ] The first step involves construction method from \cite{iqi_3}, as explained below. \\
	Jean et. al., \cite{iqi_3}  define a bijection $T$ as $T$: ($S_{2^4})^2$$\rightarrow$ $(S_{2^4})^2$. Basically, the finite field  $GF(2^8)$ is constructed  with the help of quotient structure $F_{2^8}$=$F_{2^4}[X]/P(X)$ where $P(X)=X^2+ \alpha X +\beta$, with $\alpha$, $\beta$ $\in$ $F_{2^4}$,  is an irreducible polynomial of degree 2. According to Jean et. al., \cite{iqi_3}, the elements of $F_{2^8}$ viewed as $F_{2^4}[X]/P(X)$ are polynomials of degree 1 of the form $aX+b$, denoted as couple $(a,b) \in (F_{2^4})^2$.
\begin{theorem}  Let $P(X)=X^2+\alpha X+\beta \in F_{2^4}[X]$, $P$ is irreducible if and only if $\forall$ $\gamma$ $\in$ $F_{2^4}$, $[(\alpha-\gamma)\gamma-\beta]\neq0$ \cite{iqi_3}.
\end{theorem}
definition:
\begin{definition} 
Let $P(X)=X^2+ \alpha X +\beta \in S_{2^4}[X]$, $P$ is pseudo-irreducible if and only if $\forall$ $\gamma$ $\in$ $F_{2^4}$, $[(\alpha-\gamma)\gamma-\beta]\neq0$ [3].
\end{definition}
\begin{theorem}  Let $P(X)=X^2+\alpha X+\beta \in S_{2^4}[X]$, be a pseudo-irreducible polynomial. The transformation:
\begin{center}
$T$: $(S_{2^4})^2$ $\rightarrow$ $(S_{2^4})^2$ \\
$(0,0)\rightarrow (0,0)$ \\
$(0,b)\rightarrow (0,b^{-1})$ \\
$($a$,$b$)\rightarrow (a^{-1} c,a^{-1} d)$
\end{center}
such that $\gamma=a^{-1}b$, $c=[(\alpha-\gamma)\gamma-\beta]^{-1}$, and $d=c(\alpha-\gamma)$, is a bijection.
\end{theorem}
The bijection $T$ will give us semifield S-box shown in Table 3.
\end{description}
\begin{table*}[h]
\centering 
\resizebox*{.95\textwidth}{!}{
\begin{tabular}{c | c c c c c c c c c c c c c c c c} 
\hline\hline
Rows/Columns & 0 & 1 & 2 & 3 & 4 & 5 & 6 & 7 & 8 & 9 & 10 & 11 & 12 & 13 & 14 & 15  \\
\hline\hline 
0&63&32&154&249&92&67&216&164&187&125&30&133&199&98&230&1\\
1&140&185&128&57&161&156&206&166&44&151&93&157&198&163&79&111\\
2&91&170&222&97&171&50&36&34&158&61&76&202&123&229&101&214\\
3&180&191&75&53&251&182&107&80&83&5&146&243&228&78&41&51\\
4&208&64&74&188&212&69&73&16&224&183&108&143&196&9&130&8\\
5&99&219&127&241&227&82&19&42&40&96&95&248&236&235&46&194\\
6&94&37&4&65&105&149&114&52&117&77&49&172&38&240&178&131\\
7&2&10&132&90&87&134&255&31&48&20&54&136&210&215&112&116\\
8&177&6&211&152&135&142&56&119&153&150&138&103&70&109&245&29\\
9&58&27&55&238&59&129&225&223&209&147&204&145&184&60&81&169\\
10&213&26&43&89&11&18&189&247&160&45&120&118&113&205&139&24\\
11&232&17&173&190&226&126&0&168&203&155&250&88&159&239&246&148\\
12&237&39&186&15&47&13&12&84&33&115&176&25&244&141&200&110\\
13&137&72&197&35&100&71&124&22&193&253&231&207&234&21&218&167\\
14&7&233&195&68&162&14&121&122&62&144&106&252&165&86&179&221\\
15&102&201&220&181&174&175&104&242&23&66&85&217&3&192&28&254\\
\hline 
\end{tabular}
}
\caption{ S-box based on Semifields Pseudo-extensions by \cite{iqi_3}.}
\label{ti_2} 
\end{table*}
\begin{description}
\item[Step 2: ] In this step,  the group action of $S_8$ is applied on semifield S-box of Step 1. The semifield S-box behaves like a set.
The group action leads to generation of 40320 new S-boxes, because the order of $S_8$ is 40320 and corresponding to every single permutation we have a novel S-box.  The process of synthesis is explained in Fig.\hspace{1mm}\ref{fig_iqi_1} and mathematical expression is given below:

The action \(S_8 \ S-boxes: \ S_8  \times Semifield \ S-box \ \rightarrow   Semifield \ S-boxes\)

	defined as $$\mu (a_0 a_1 a_2 a_3 a_4 a_5 a_6 a_7) \ = \ (a'_0 a'_1 a'_2 a'_3 a'_4 a'_5 a'_6 a'_7)$$  where,
\( \mu\in S_8\) and,   $$\  (a_0 a_1 a_2 a_3 a_4 a_5 a_6 a_7), (a'_0 a'_1 a'_2 a'_3 a'_4 a'_5 a'_6 a'_7) \in Semifield-S-box$$.
\end{description}
\begin{figure*}[!htb]
\centering
\includegraphics[width=130mm]{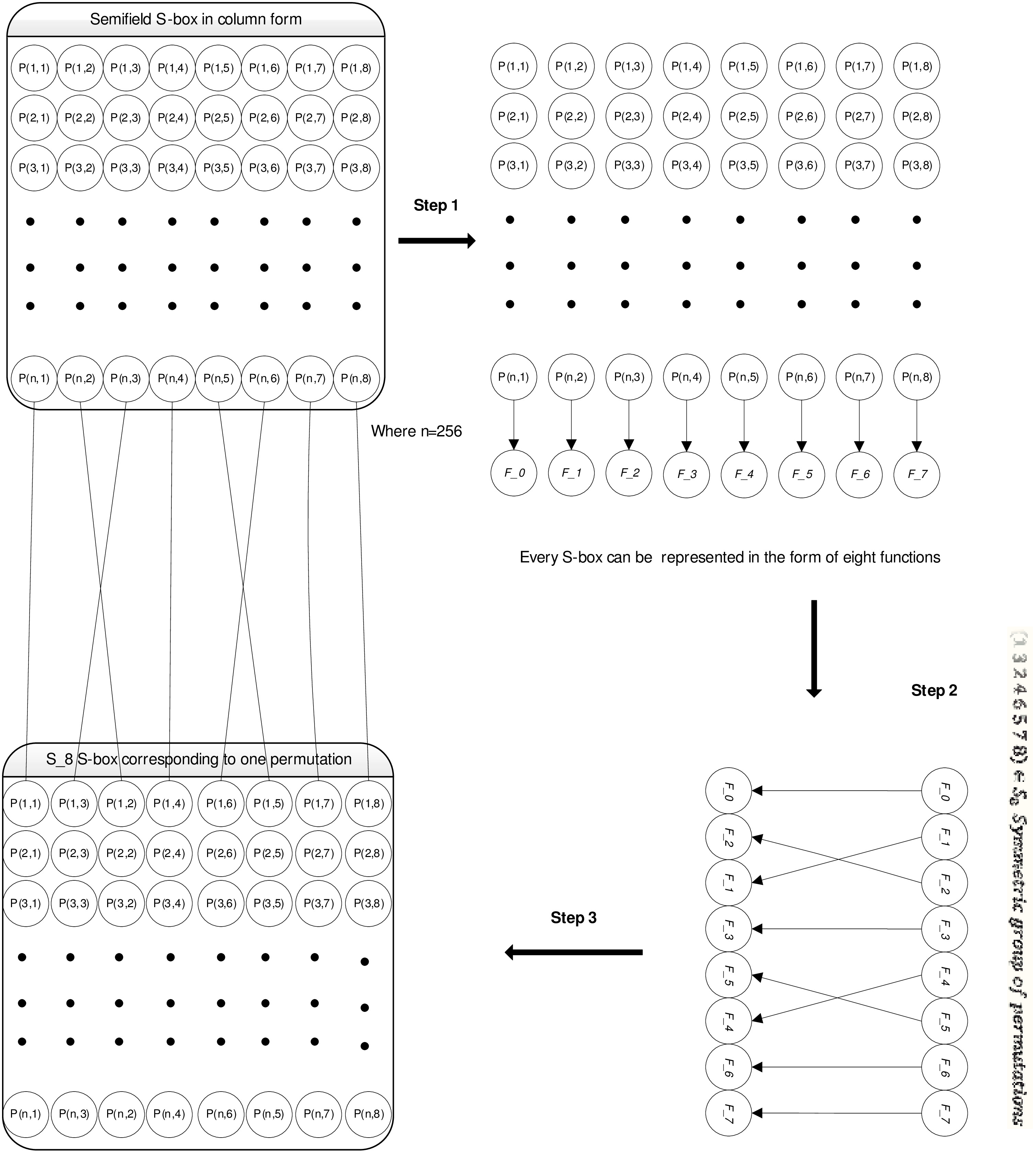}
\caption{Graphical illustration of proposed algorithm}
\label{fig_iqi_1}
\end{figure*}

Fig.\hspace{1mm}\ref{fig_iqi_1}, explains the mathematical action in a simplified way with three steps. Initially, in Step 1, we have converted the elements of semifield S-box from decimal to binary values. So we have eight different Boolean functions $F_0, F_1,F_2,F_3,F_4,F_5,F_6,F_7$ as shown in Fig.\hspace{1mm}\ref{fig_iqi_1}. In step 2 we have taken an arbitrary permutation $(1 3 2 4 6 5 7 8)$ from $S_8$ for the construction of one S-box. On applying this permutation on $F_0, F_1,F_2,F_3,F_4,$ $F_5,F_6,F_7$, the effect of permutation on Boolean functions will give us a new arrangement which is as follows: $F_0, F_2,F_1,F_3,F_5,F_4,F_6,F_7$. At the end the binary number is converted once again in the form of decimal values get new $S_8$ semifield S-box.
\subsection{Cryptographic strength analyses of S-box}

To measure the good properties of S-box benchmark criteria are presented in literature like bijectivity, differential approximation probability (DP), strict avalanche criterion (SAC), bit independence criterion (BIC), nonlinearity, and linear probability (LP).
We carry out the security analysis of the proposed S-box of example given in Table\hspace{1mm}\ref{ti_2} using these well known criteria.

\begin{enumerate}
	\item Bijectivity: \cite{iqi_3} If the linear sum of the Boolean function $f_i$ of each component of the designed $n \times n$ S-box is $2^{n-1}$  , then $f$ is a bijection . Mathematically, we can write as

\begin{equation}
	wt(a_1f_1+a_2f_2+...+a_n f_n).
\end{equation}

where $a_i\in{0,1}$, $(a_1,a_2,...,a_n)\neq(0,0,...,0)$, $wt()$ denotes the Hamming weight. In effect, an inverse is important specifically in a substitution netwrok, therefore S-box should be bijective.

\item Nonlinearity: The nonlinearity of an S-box can be tested by the following formula:

\begin{equation}
	N_f=2^{-n}\left(1-max_{\omega \in GF(2^n)}\left|2^{-n}\sum_{x\in GF(2^n)} (-1)^{f(x)\oplus x.\omega}\right|\right),
\end{equation}

where $\omega \in GF(2^8)$.

\item Strict avalanche criterion: This analysis depicts information that while one bit of eight lengths input byte of plaintext modifies, will yield a 0.5 probability of the outcomes changes in byte of 8 bits balanced for entries.

\item Bit independent criterion: For two Boolean function $f_j$, $f_k$, one can test the independence criterion of a substitution box by validating if,  for any two output bits of the S-box, $f_j \oplus f_k$ $(j \neq k)$ fulfills the SAC and nonlinearity.

\item XOR table and differential invariant: XOR table of substitution box basically depends on the calculation of $\rho_{L}(a,b)= \left\{ x\in GF(2^8): L(x)\oplus L(a\oplus x)= M\right\}$ $\forall a,b \in GF(2^8)$. The differential invariant $\rho_{L}(a,b)$ is found as follows:

$\rho_{L}(a,b) =\underbrace{max}_{a,b \in GF(2^8), a \neq 0}\left|\left\{ x\in GF(2^8): L(x)\oplus L(a\oplus x)= M\right\}\right|$. \\

\end{enumerate}
\subsection{Experimental results for proposed S-Boxes}

\begin{table*}[h]
\centering 
\resizebox*{.95\textwidth}{!}{
\begin{tabular}{l c c c c c c } 
\hline\hline
S-boxes & Non-linearity & SAC &	BIC of Non-linearity &	BIC of SAC & DP & LP\\
\hline\hline 
AES S-box \cite{aes}							&	112	&	0.504	&112	&	0.504	&	0.015625	&0.0625\\
Semifield S-box \cite{iqi_3}				&	112	&	0.503	&	112	&	0.501	&	0.015625	&	0.0625 \\
$S_8$ Semifield S-boxes		&	112	&	0.503	&	112	&	0.501	&	0.015625	&	0.0625\\
\hline 
\end{tabular}
}
\caption{ Comparitive analysis of proposed $40320$ $S_8$ semifield S-boxes with prevailing S-boxes}
\label{ti_3} 
\end{table*}
In Table\hspace{1mm}\ref{ti_3}, we have shown some analyses such as linear approximation probability, differential approximation probability, strict avalanche criterion, average bit independence criterion for nonlinearity, average bit independence criterion for strict avalanche criterion and nonlinearity of proposed $S_8$ semi-field S-boxes. Further, we have compared the cryptographic strength of presented 40320 boxes with semi-field S-box of \cite{iqi_3} and advanced encryption standard (AES) S-box \cite{aes}. It can be observed that all the properties of newly generated nonlinear components are almost equivalent to S-box of \cite{iqi_3} and \cite{aes}. Hence, we can say that proposed 40320 S-boxes are a powerful as AES S-box, which is known as one of the most invulnerable S-box against all kinds of differential and linear attacks.
\section{Proposed lightweight block cipher}

\begin{figure*}
\centering
\includegraphics[width=150mm]{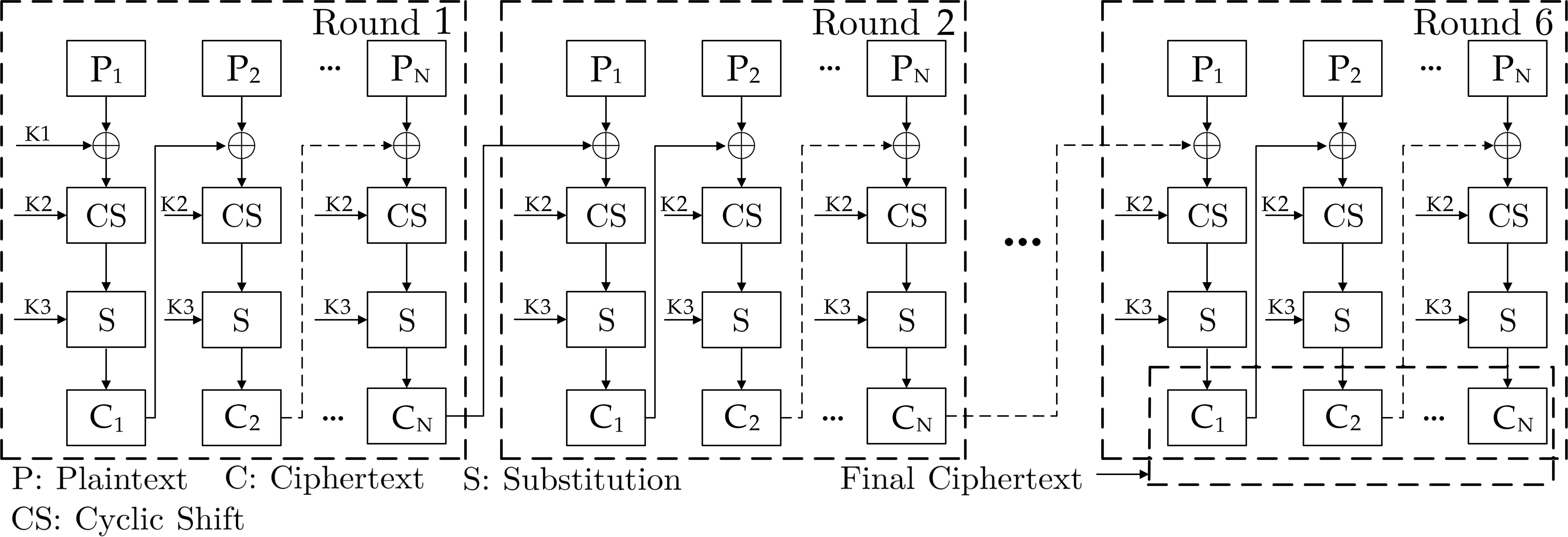}
\caption{A top level block description of proposed lightweight block cipher. The plaintext in blocks is encrypted through substitution-permutation (cyclic shift and substitution) network in six rounds to get the ciphertext.}
\label{encr_algo}
\end{figure*}

For the application of proposed S-boxes presented in the previous section, we have proposed a lightweight block cipher which utilizes these S-boxes and can be employed for low profile applications. The proposed cipher consists of basic confusion-diffusion network performed by substitution and permutation. The permutation is performed with the help of cyclic shift operation and substitution is done with the help of numerous proposed S-boxes having same algebraic properties. The top level block description of proposed lightweight block cipher is illustrated in Fig.\hspace{1mm}\ref{encr_algo}.

The plaintext is divided and considered as individual blocks of 128-binary bits, let the representation of 128-binary bits block of plaintext be \(B_p\). There is whitening step beside the confusion-diffusion network which is performed at the very beginning. The first block of plaintext is xored with the user supplied 128-binary bits key, \(K_1\). After the xor operation (whitening), the xored block which is represented as \(B_x\) is the input to the substitution and permutation modules. The next blocks of plaintext are xored with the immediate previous blocks of ciphertext.
\begin{algorithm}
\caption{The 6-round encryption process of proposed block cipher.}\label{euclid}
\vspace{1mm}
\textbf{Inputs:} Plaintext (blocks form) \(B_p\), User provided 128-bit key, \(K_1\), secret initial values of modified logistic map, \(K_2\) and TD-ERCS map, \(K_3\), S-Boxes \(S_{1 \times S_n}\).

\textbf{Output:} Ciphertext (blocks form), \(C_p\).

\begin{algorithmic}[1]
\For {\(j \leftarrow 1:6\)}
	\For{\(i \leftarrow 1:no\_of\_blocks\)}
			\State \(B_x \leftarrow\) \(bitxor(B_p,K_(1))\)
			\State \(B_c \leftarrow\) \(cyclic shift(B_x,K_2)\)
			\State \(C_p \leftarrow\) \(substitution(B_c, S(K_3))\)
			\State K\((1) = C_p\)
	\EndFor
	\State \(\textbf{Endfor}\)
\EndFor
\State \(\textbf{Endfor}\)
\end{algorithmic}
\end{algorithm}
As mentioned earlier, binary cyclic shift is performed for the permutation. Let us take \(x\) to be the chaotic sequence obtained from modified logistic chaotic map for the initial values \(x_0\), \(b\) and \(r\). These initial values are considered as the first three secret keys of the proposed encryption algorithm which combines to give \(K_2\) as illustrated in Fig.\hspace{1mm}\ref{encr}, that is, \(k_2^{1} = x_0\), \(k_2^{2} = b\) and \(k_2^{3} = r\). The length of \(x\) is 128 which has values under modulo 8.
Since the initial range of chaotic sequence is [0, 1], therefore we multiply that sequence with 100 to amplify the range and then limit the the sequence under modulo 8.
 In parallel, let the binary representation of first 8 bits of xor block be \(P_x\). For each \(P_x\), there is a left cyclic shift of \(x_{q}\) binary bits which results in a cyclic shifted 8 bits, \(P_c\) and correspondingly cyclic shifted block, \(B_c\). For instance, let \(P_x = 11010110\) and \(x_{q} = 3\), then after cyclic shift, \(P_c = 10110110\).

Take \(x\) and \(k\) be the two chaotic sequences obtained from TD-ERCS chaotic map having initial values \(x_0\), \(tan\alpha\), \(\mu\) and \(m\). For simplicity and convenience,  \(x\) is denoted as \(y\). These four initial values are considered as next four secret keys of the proposed encryption algorithm, that is, \(k_3^{1} = x_0\), \(k_3^{2} = tan\alpha\), \(k_3^{3} = \mu\) and \(k_3^{4} = a\) which combine to give \(K_3\). Since, we need only one chaotic sequence from this map, therefore we will only use \(y\) sequence.
The sequence  \(y\) has length 16 and values under modulo \(S_n\), where, \(S_n\) denotes the total number of S-boxes used.
Initially the chaotic sequence  has the range of [-1, 1], so we amplify that range by first shifting the range from [-1, 1] to [0, 2] and then multiplying that shifted sequence with \(1000\), and finally limiting the sequence under modulo \(S_n\). In parallel, for the substitution step, combinations of 8 bits in \(B_c\) are substituted with the 8 bits of one of the proposed S-boxes. The decision of choosing which S-box to utilize is taken by the chaotic sequence of \(y\). Let the binary representation of first 8 bits of \(B_c\) be \(P_c\). For each \(P_c\), the eight bits are split into 4 Most Significant Bits (MSBs) and 4 Least Significant Bits (LSBs) first. These MSBs and LSBs are then converted into decimal and these decimal values corresponds to the row and column numbers of a particular S-box. The element at that position of specific S-box is converted into binary having 8 bits and will be substituted in place of \(P_c\). For instance, let \(P_c = 10110110\) and \(y_{q} = 105\), then the 4 MSBs are 1011 and 4 LSBs are 0110, the decimal value at 11th row and 5th column of 105th S-box will be substituted to get the 8 substituted binary bits \(P_s\) and correspondingly ciphertext block, \(C_p\).

The proposed block cipher in form of pseudo-code is illustrated in Algorithm 1. It has only six rounds as contrary to other renown block ciphers which have more than 12 rounds. Despite few rounds, the results are very competitive to the benchmark algorithms due to the multiple S-boxes employed instead of a single S-box. The experimental, statistical and security analysis are presented in the next sections to illustrate the robustness and performance effectiveness of the proposed block cipher.
\section{Experimental results and statistical analysis}

\begin{figure}
  \centering
		\subfloat[]{\includegraphics[width=0.25\textwidth]{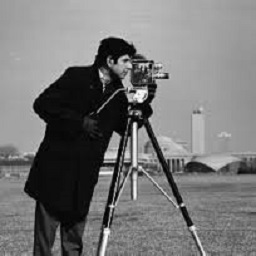} }
	  \subfloat[]{\includegraphics[width=0.25\textwidth]{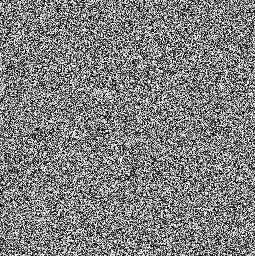} } \\
	  \subfloat[]{\includegraphics[width=0.25\textwidth]{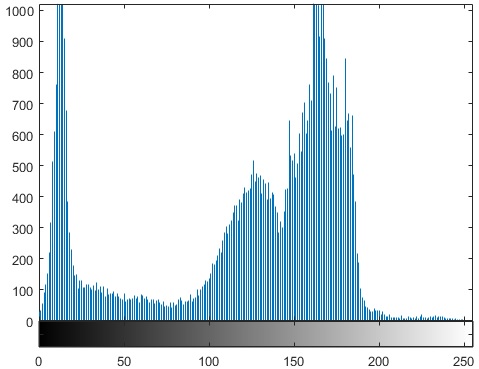} }
		\subfloat[]{\includegraphics[width=0.25\textwidth]{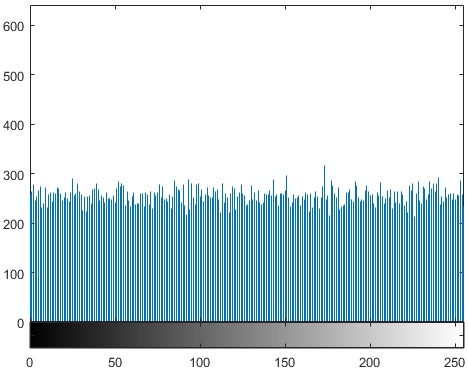} }
  \caption{(a) Cameraman image having size of $256\time256$ is considered as plaintext. (b) Encrypted cameraman image when the proposed block cipher is applied on the original cameraman image, it can be observe that it is completely random and gives no information regarding the plainimage. (c) Histogram of cameraman image and (d) histogram of encrypted cameraman image.}
	\label{encr}
\end{figure}

\begin{figure}
  \centering
		\subfloat[]{\includegraphics[width=0.25\textwidth]{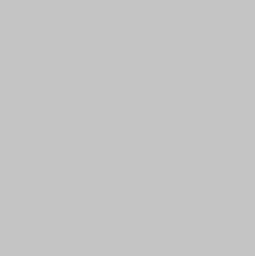} }
	  \subfloat[]{\includegraphics[width=0.25\textwidth]{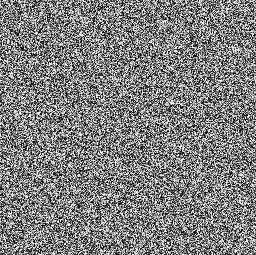} } \\
	  \subfloat[]{\includegraphics[width=0.25\textwidth]{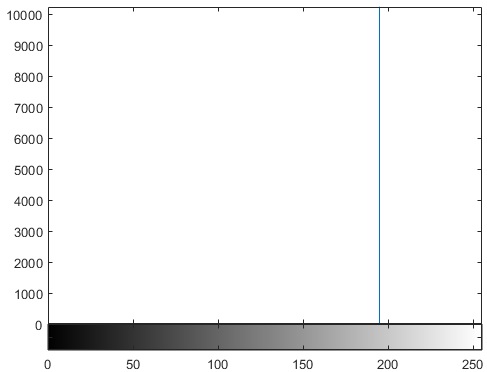} }
		\subfloat[]{\includegraphics[width=0.25\textwidth]{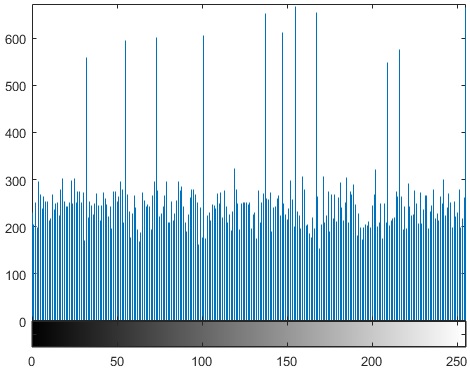} }
  \caption{(a) One-scale image which has only one gray value having perfect auto-correlation. (b) The encrypted version of this one-scale image (c) Histogram of one-scale and (d) histogram of its encrypted version.}
	\label{onescale}
\end{figure}

For experiments, we have taken two dimensional digital images as plaintext. First, the cameraman image having size of $256\time256$ is considered as plaintext shown in Fig.\hspace{1mm}\ref{encr}(a). This plainimage is then encrypted using proposed lightweight block cipher and encrypted cameraman image is shown in Fig.\hspace{1mm}\ref{encr}(b) which is completely random and gives no information regarding the original plainimage. To see the distribution of values (image pixels) in plainimage and cipherimage, we have plotted the histograms of these images shown in Fig.\hspace{1mm}\ref{encr}(c) and Fig.\hspace{1mm}\ref{encr}(d) respectively. It can be observed that the distribution of the values of cipherimage is uniform showing good resistance against frequency analysis. We have also considered a one-scale image which has only one gray value having perfect auto-correlation as shown in Fig.\hspace{1mm}\ref{onescale}(a).  The encrypted version of this one-scale image is shown in Fig.\hspace{1mm}\ref{onescale}(b). The histogram of one-scale and its encrypted version is plotted and shown in Fig.\hspace{1mm}\ref{onescale}(c) and Fig.\hspace{1mm}\ref{onescale}(d) respectively. It can be seen that the histogram of the encrypted version is flat again, despite maximum auto-correlation in the plainimage.

\subsection{Statistical analysis}

For further examination of the visual strength of the proposed block cipher, we have performed different statistical analysis and have carried out a comparison of our results with the state-of-art block cipher. We have considered the following statistical analyses.

\subsubsection{Correlation}

For an image, the correlation is defined as \cite{stat}:

\begin{equation}
Corr. = \sum_{a, b} \frac{(a - \mu a)(b - \mu b)\rho(a, b)}{\varphi_{a}\varphi_{b}}
\end{equation}

where \(a, b\) refer to position of image pixels, \(\rho(a, b)\) represents the pixel value at \(a^{th}\) row and \(b^{th}\) column of image matrix, $\mu$ and $\varphi$ respectively denote the variance and the standard deviation. The correlation analysis measures the extent to which two neighboring image pixels are similar, over the whole image. It has the range \([-1 \ 1]\) where the correlation value 1 means the ideal correlation.

\subsubsection{Entropy}

The image entropy is defined as \cite{stat}:

\begin{equation}
Entropy = -\sum_{a, b} pr(\rho(a, b)) \log_{2} pr(\rho(a, b)).
\end{equation}

where \(a, b\) refer to position of image pixels, \(\rho(a, b)\) represents the pixel value at \(a^{th}\) row and \(b^{th}\) column of image matrix and \(pr(\rho(a, b))\) is the probability of image pixel.  Entropy depicts the uncertainty of the image and has range \([0 \ 8]\) for an image with 256 gray scales. The more the amount of entropy is, the more the amount of uncertainty is.

\subsubsection{Contrast}

The image contrast is defined as \cite{stat}:

\begin{equation}
Contrast = \sum_{a, b} |a - b|^{2} \rho(a, b).
\end{equation}

where \(a, b\) refer to position of image pixels, \(\rho(a, b)\) represents the pixel value at \(a^{th}\) row and \(b^{th}\) column of image mtrix. The contrast analysis of the image is carried out for the purpose of identification of objects in texture of an image.
 The range of contrast values is \([0 \ (size(Image)-1)^2]\). For a constant image, the contrast value is 0. The more the amount of contrast is, the more is the variation in image pixels.

\subsubsection{Homogeneity}

The image homogeneity is defined as \cite{stat}:

\begin{equation}
Homo. = \sum_{a, b}\frac{\rho(a, b)}{1 + |a - b|}
\end{equation}

where \(a, b\) refer to position of image pixels. The nearness  of the distribution in the gray level cooccurrence matrix (GLCM) to GLCM diagonal is determined by homogeneity analysis. The homogeneity values range is \([0 \ 1]\).

\subsubsection{Energy}

The image energy is defined as \cite{stat}:

\begin{equation}
Energy = \sum_{a, b} \rho(a, b)^{2}.
\end{equation}

where \(a, b\) refer to position of image pixels. The energy analysis provides the sum of squared elements in the GLCM. The energy values have range \([0 \ 1]\). For a constant image, the energy value is 1.

Table\hspace{1mm}\ref{t_3} shows the outcome of the above analyses for the proposed image encryption scheme. It also provides a comparative study with the benchmark blockcipher of AES. It is worth noticing that the presented scheme depicts competitive results despite  significant lower computational complexity.
\begin{table}[h]
\centering 
\begin{tabular}{p{1.1cm} | p{1cm} p{1cm} p{1cm} p{1cm} p{1cm}} 
\hline\hline
Analysis & Corr. & Entropy & Homo. & Contrast & Energy  \\
\hline\hline 
Prop. & -4e-4 & 7.9974 & 0.9952 & 0.0434 & 0.9911       \\
AES & -4e-3 & 7.9967 & 0.9934 & 0.3678 & 0.9851       \\

\hline 
\end{tabular}
\caption{Comparative statistical analysis of proposed and AES cipher.} 
\label{t_3} 
\end{table}

\section{Security analysis}

For the robustness and to check whether the proposed cipher can resist against known attacks, we have carried out series of security analysis that will ensure the effectiveness of proposed cipher.

\subsection{Key space and key sensitivity}

The  total number of secure keys that can be used in the cipher is referred as key space. In our proposed work, we have used three keys, in which, the first one is the user provided 128 binary bit key, second and third are the combination of initial values of the two chaotic maps used. The total combination of these three secure keys is sufficiently large so, for checking all the combinations, any modern computer should  take more than $10^{10}$ years.  Thus the proposed cipher can resist to any brute force attack.

\begin{figure}
  \centering
		\subfloat[]{\includegraphics[width=0.25\textwidth]{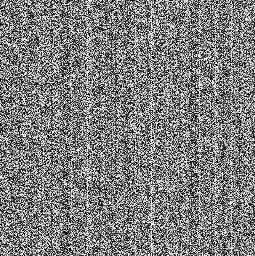} }
	  \subfloat[]{\includegraphics[width=0.25\textwidth]{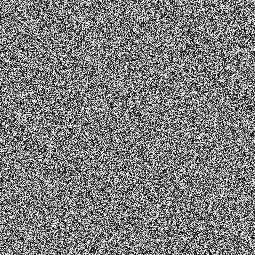} } \\
	  \subfloat[]{\includegraphics[width=0.25\textwidth]{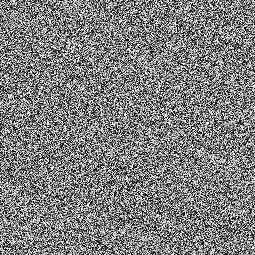} }
		\subfloat[]{\includegraphics[width=0.25\textwidth]{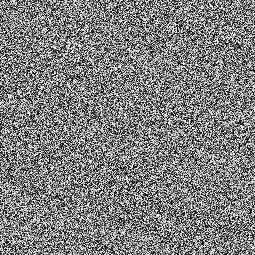} }
  \caption{Decrypted image in which \(K_1\) is changed by a single binary bit. Decrypted image for which, the decryption key of \(b\) of \(K_2\) is changed from \(b = 0.2000000000\) to \(b = 0.2000000001\). Decrypted for which, the decryption key of \(r\) of \(K_2\) is changed from \(r = 4.5000000000\) to \(r = 4.5000000001\). Decrypted for which, the decryption key of \(x_0\) of \(K_3\) is changed from \(x_0 = 0.5000000000\) to \(x_0 = 0.5000000001\). }
	\label{ses}
\end{figure}

However, key space alone is not enough to guarantee the effectiveness of a cipher in this regard. It is necessary that with the key space, key sensitivity should be achieved as well. Key sensitivity refers to a property of cipher in which the decryption of ciphertext can not be successively achieved despite the fact that there is a minor change in the encryption and decryption keys. For the analysis, first, we have taken the cameraman image (shown in Fig.\hspace{1mm}\ref{encr}(a)) and encrypted with the proposed cipher (Fig.\hspace{1mm}\ref{encr}(b) shows the encrypted image). For the decryption, we have slightly changed the decryption keys and attempted decrypting the encrypted images with these slightly altered decryption keys. Fig.\hspace{1mm}\ref{ses}(a) shows the decrypted image in which \(K_1\) is changed by a single binary bit. Fig.\hspace{1mm}\ref{ses}(b) shows the decrypted image in which \(b\) of \(K_2\) is changed from \(b = 0.2000000000\) to \(b = 0.2000000001\). Fig.\hspace{1mm}\ref{ses}(c) shows the decrypted image in which \(r\) of \(K_2\) is changed from \(r = 4.5000000000\) to \(r = 4.5000000001\). Fig.\hspace{1mm}\ref{ses}(d) shows the decrypted image in which \(x_0\) of \(K_3\) is changed from \(x_0 = 0.5000000000\) to \(x_0 = 0.5000000001\). It can be seen that, even for a small modification in the decryption keys, decryption is unsuccessful in all of considered images.

\subsection{Avalanche analysis}

The avalanche effect refers to an appealing characteristics of cryptosystems. The avalanche effect is evident when a slight change in the input (for instance, flipping a single bit) leads to a notable change in the output (like change in half the output bits). Mathematically,   the avalanche effect can be determined by the following formulas for the number of pixel change rate (NPCR) and the unified average change intensity (UACI) \cite{npcr_uaci}.

\begin{equation}
	NPCR = \frac{\sum_{a,b}D(a,b)}{N \times M}\times100 \%,
\end{equation}
\begin{equation}
	UACI = \frac{1}{N \times M}\left[\sum_{a,b}\frac{|A_1(a,b) - A_2(a,b)|}{255}\right] \times 100 \%,
\end{equation}

where \(A_1\) and \(A_2\) are the two encrypted images acquired from the original image with the disparity of a single pixel, \(N\) and \(M\) represent the number of rows and columns of encrypted image matrix,  and \(D(a,b)\) is defined as
	\[ D(a,b) = \left\{
  \begin{array}{l l}
    0 & \quad \text{if \(A_1(a,b) = A_2(a,b)\)},\\
    1 & \quad \text{if \(A_1(a,b) \neq A_2(a,b)\)}.
  \end{array} \right.\]

	NPCR yields the rate of change  of number of pixels of encrypted image once a single pixel of original image is amended.
	UACI determines  normal power of contrasts between original and encrypted images.
	The least required value for NPCR must be 50\%. NPCR and UACI analysis is carried out for some standard images of image processing such as Lena, cameraman and baboon.
	We have taken three cases to analyze the NPCR and UACI cryptographic strength for propoased image encryption technique, for each of the considered images.
	For analysis, firstly a pixel is chosen in the first two rows of the original image and then the procedure is repeated for two other parts (roughly in middle and end) of the data. The comparative analysis of the proposed image encryption technique with the AES for NPCR and UACI is provided in Table\hspace{1mm}\ref{t_npcr}. It can be seen that proposed algorithm is competing with the most widely applicable and economical block cipher cryptosystem.

	\begin{table}[h]
\centering
  \begin{tabular}{p{.9cm} p{1.1cm} p{1cm} p{1cm} p{1cm} p{1cm}}
    \hline \hline
      \multicolumn{2}{c}{Analysis} &
      \multicolumn{2}{c}{NPCR(\%)} &
      \multicolumn{2}{c}{UACI(\%)} \\
			\hline
			\multicolumn{2}{c}{Images \& Loc.} & Prop. & AES & Prop. & AES \\
			\hline \hline
    \multirow{3}{*}{Cman} & first & 99.5463 & 99.6048 & 33.1521 & 33.5360 \\
		& mid & 99.5412 & 99.6201 & 33.2512 & 33.5212 \\
		& last & 99.5874 & 99.5819 & 33.3981 & 33.5245 \\
    \hline
		\multirow{3}{*}{Lena} & first & 99.2632 & 99.6094 & 33.4521 & 33.3996 \\
		& mid & 99.3698 & 99.6506 & 33.3874 & 33.3139 \\
		& last & 99.4529 & 99.6002 & 33.5214 & 33.5133 \\
    \hline
		\multirow{3}{*}{Baboon} & first & 99.4563 & 99.6124 & 33.5210 & 33.4463 \\
		& mid & 99.5210 & 99.6033 & 33.2363 & 33.4561 \\
		& last & 99.5632 & 99.6185 & 33.3152 & 33.5252 \\
    \hline
		\multirow{2}{*}{Key S.} & Case I & 99.5298 & 99.5972 & 33.6210 & 33.5029 \\
		& Case II & 99.6210 & 99.6460 & 33.3698 & 33.5468 \\
    \hline
  \end{tabular}
\caption{Comparative NPCR and UACI analysis of proposed cipher with AES on the images of cameraman, lena and baboon for three cases: a single bit change in first pixel, in mid pixel and in last pixel. The analysis of two different cases for key sensitivity are also presented here.} 
	\label{t_npcr}
\end{table}

Moreover, we have done NPCR and UACI analysis to analyze the key security of proposed key schedule. We consider following two cases.
\begin{description}
\item [Case I] We have applied two different encryption keys having a single bit difference on same original image to get two significantly different encrypted images having NPRC and UACI values 99.5298 and 33.6210 respectively.
\item [Case II] Here we have applied two different
decryption keys having a single bit difference on same encrypted image to get two significantly different decrypted images having NPRC and UACI values 99.6210 and 33.3698 respectively.
\end{description}
The comparative analysis of the proposed image encryption technique with the AES for key sensitivity using NPCR and UACI is given in Table\hspace{1mm}\ref{t_npcr}. The values of the analysis show that the proposed key schedule is invulnerable for image encryption.
\subsection{Cryptanalysis}

To analyze the full cryptographic strength of proposed image cryptosystem against known attacks, it is required to pass plainimage through complete rounds. The known attacks are as follows.

\subsubsection{Linear cryptanalysis}

Th linear approximation probability (LP) is utilized to measure the discrepancy of an event. This test is useful in determining the optimum value of discrepancy in the outcome of event.  The two masks, $\Gamma y$ and $\Gamma x$, are utilized for parity of the output and input bits, respectively.
For the nonlinear component of block cipher, LP is defined as \cite{linear_crypt}:

\begin{equation}
	LP = max_{\Gamma x \Gamma y \neq 0}\left|\frac{\#\left\{x/X \bullet \Gamma x = S(x)\bullet \Gamma y = \Delta y\right\}}{2^{n}} - \frac{1}{2}\right|,
\end{equation}

where all possible unique inputs are contained in the set  \(X\) and the cardinality of \(X\) is \(2^{n}\).
The average maximum value of  \(LP\) of proposed cryptosystem having 1 round and 8 S-boxes is \( LP_{max} = 2^{-4.21}\).
Furthermore, the generalization of proposed cryptosystem with 256 active S-Boxes and  4-rounds will give us average maximum value \(LP_{max}^{4r} = 2^{-4.21 \times 256} = 2^{-1077}\). It is worth mentioning that to a certain extent, it is difficult to launch linear cryptanalysis against proposed cryptosystem.

\subsubsection{Differential cryptanalysis}

With a specific end goal to guarantee uniform mapping, the differential at input should exceptionally guide to an output differential. These attributes ensure uniform mapping probability for each input bit \(i\). The differential approximation probability scheme for S-Box determines  the differential uniformity, defined as \cite{diff_crypt}:

\begin{equation}
	DP(\Delta x\rightarrow \Delta y) = \left[\frac{\#\left\{x\in X /S(x)\oplus S(x\oplus \Delta x) = \Delta y\right\}}{2^{m}}\right],
\end{equation}

where $\Delta y$ and $\Delta x$ are output  and input differential, respectively.
We have applied 256 S-boxes with good cryptographic properties and have extended proposed algorithm for 4 rounds to get good results against \(DP\). \(DP_{max} = 2^{-4.05}\) is the average maximum \(DP\) of the S-Boxes used. The value of \(DP\) of the proposed cryptosystem guarantees the cryptographic strength of the proposed algorithm against differential cryptanalysis.


%

\section{Conclusion}

This paper is concerned with the study of S-boxes based on semifield  and their application to develop a block cipher that can be employed for low profile applications.
Although the currently known S-boxes based on finite fields perform well against known algebraic attacks, it is possible that there could be some unknown algebraic attacks that may beat the confusion creating capability of finite field S-boxes.
In \cite{iqi_3}, it was proposed to consider finite semifields instead of finite fields to generate S-Boxes in order to to enhance the algebraic strength of S-boxes against unknown algebraic attacks.
The algebraic structure of semifield is not commonly used in block cipher cryptography because it does not have complete classification up to finite order. In fact, the known classification of semifields is up to $2^6$. This makes it challenging to use the small structure $2^4$ for the construction of $2^8$ S-boxes. Dumas- Orfila \cite{iqi_3} utilized pseudo-extensions of semifield of order $2^4$ to construct 12781 non-equivalent S-Boxes with good  cryptographic properties. Here we establish an effective procedure for generating \(S_8\) semifield S-boxes having same algebraic properties. Utilizing the action of symmetric group of permutation on every non-equivalent S-box of \cite{iqi_3}, corresponding to each and every non-equivalent S-box of Dumas- Orfila \cite{iqi_3}, we generate $8!$ S-boxes with same cryptographic properties.
In total, this leads to 515329920 new S-boxes based on semifield and symmetric group.

As application, a block cipher is proposed utilizing \(S_8\) semifield substitution boxes and two  chaotic maps. The proposed cipher consists of three modules which are operated for six rounds. These three modules are substitution, permutation and XOR whitening. In the substitution process, multiple proposed S-boxes are used to get desired confusion between the ciphertext and secret key. This module enhances security, and the desired security is attained in less number of rounds as compared to conventional block ciphers.  Each module is applied in agreement with the specific chaotic sequence generated through a distinct chaotic map. Security analysis is performed and the simulation results confirm that the proposed algorithm is secure against well known attacks.

Finally, the proposed encryption algorithm is flexible that can be adapted to changes as required. For instance, it is possible to increase or decrease the number of rounds and S-boxes as well as to replace semifield S-boxes with other S-boxes.

\end{document}